\documentclass{article}
\usepackage{spconf,amsmath,graphicx}
\usepackage[numbers,sort&compress]{natbib}

\usepackage{balance}
\usepackage{bm}
\usepackage{amsmath, amssymb}
\usepackage{booktabs}
\usepackage{subfigure}
\usepackage{marvosym}
\usepackage[hidelinks]{hyperref}

\newcommand\eg{\emph{e.g.}}
\newcommand\ie{\emph{i.e.}}

\usepackage{color}

\title{Multi-View Spectrogram Transformer for Respiratory Sound Classification}
\name{Wentao He$^{\star,\dagger}$, Yuchen Yan$^{\star,\dagger}$, Jianfeng Ren$^{\textrm{\Letter},\dagger,\ddagger}$, Ruibin Bai$^{\dagger,\ddagger}$, Xudong Jiang$^{\S}$ \thanks{$^{\star}$ The authors contributed equally. $^{\textrm{\Letter}}$ Corresponding author.} 
\address{${}^{\dagger}$School of Computer Science, University of Nottingham Ningbo China\\
$^{\ddagger}$Nottingham Ningbo China Beacons of Excellence Research and Innovation Institute, \\
University of Nottingham Ningbo China\\
$^{\S}$School of Electrical \& Electronic Engineering, Nanyang Technological University}
\thanks{This work is supported in part by the National Natural Science Foundation of China under Grant 72071116 and in part by the Ningbo Science and Technology Bureau under Grants 2019B10026 and 2022Z173.}
}

\begin{document}
%
\maketitle
\begin{abstract}
Deep neural networks have been applied to audio spectrograms for respiratory sound classification. Existing models often treat the spectrogram as a synthetic image while overlooking its physical characteristics. In this paper, a Multi-View Spectrogram Transformer (MVST) is proposed to embed different views of time-frequency characteristics into the vision transformer. Specifically, the proposed MVST splits the mel-spectrogram into different-sized patches, representing the multi-view acoustic elements of a respiratory sound. The patches and positional embeddings are fed into transformer encoders to extract the attentional information among patches through a self-attention mechanism. Finally, a gated fusion scheme is designed to automatically weigh the multi-view features to highlight the best one in a specific scenario. Experimental results on the ICBHI dataset demonstrate that the MVST significantly outperforms state-of-the-art methods for classifying respiratory sounds. The code is available at: \url{https://github.com/wentaoheunnc/MVST}.

\end{abstract}
\begin{keywords}
Respiratory sound classification, Mel-spectrogram, Vision Transformer, ICBHI dataset
\end{keywords}
\section{Introduction}
\label{sec:intro}

\begin{figure*}[!t]
    \centering
    \includegraphics[width=0.95\textwidth]{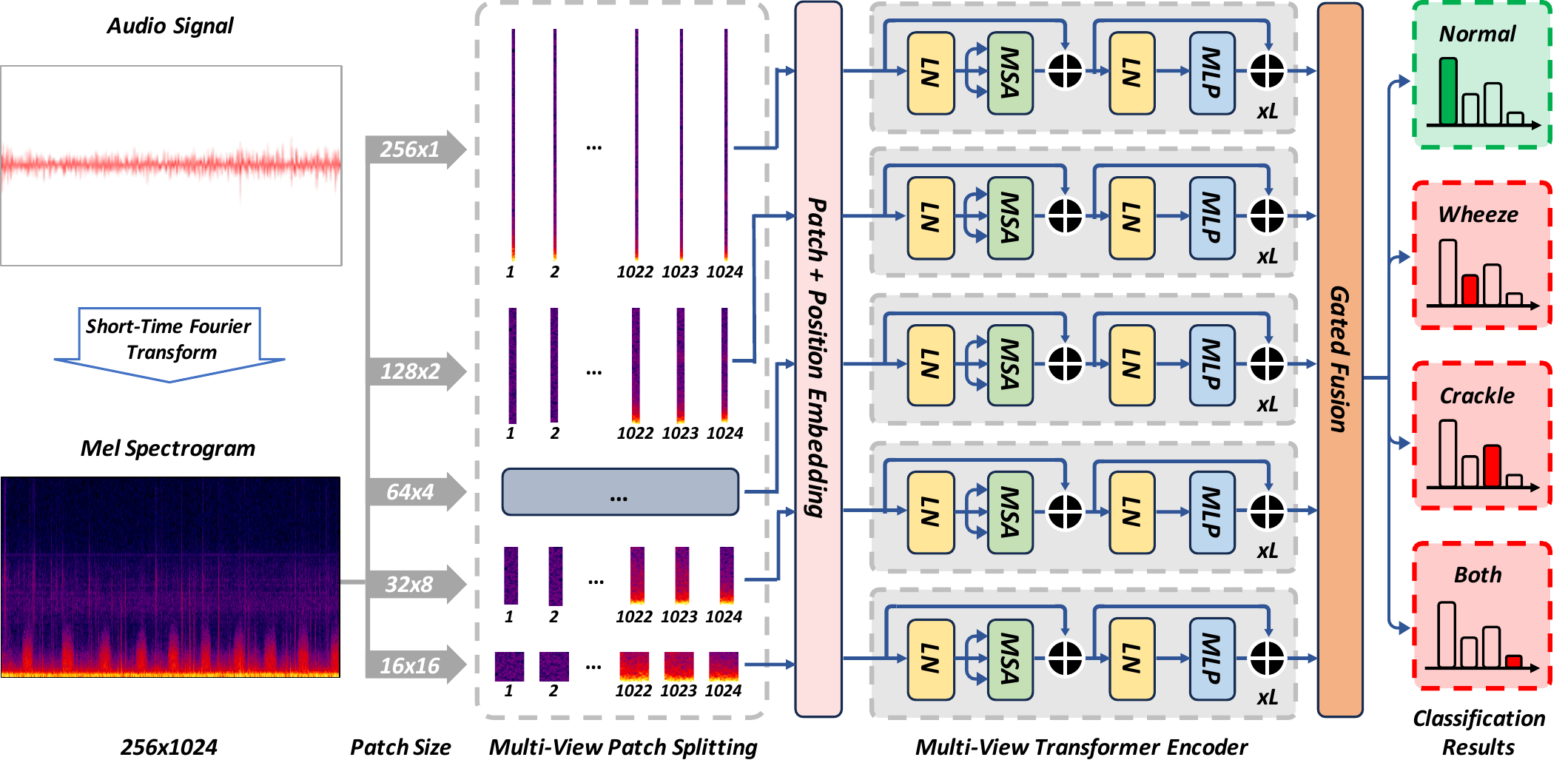}
    \caption{Overview of proposed Multi-View Spectrogram Transformer (MVST). The mel-spectrogram of an input audio is split into different-sized patches, to analyze the multi-view spectral characteristics. The multi-view spectrogram patches are then processed by multi-scale transformer encoders to capture the attentional information among patches. A gated fusion mechanism is then designed to highlight the most suitable spectral view for respiratory sound classification. 
    }
    \label{fig:main-diagram}
\end{figure*}

Respiratory disease has become one of the leading causes of death and disability worldwide. Automatic auscultation of respiratory sounds helps doctors for early diagnosis of lung diseases with adventitious breathing sounds such as crackles and/or wheezes \cite{rocha2018alpha}. Many systems have been developed for automated respiratory sound classification. One approach is to transform an audio signal to a time-frequency representation using short-time Fourier transform (STFT) \cite{nguyen2022lung,xu2021arsc,ma2019lungbrn,li2021lungattn,gairola2021respirenet,wang2022domain,zhao2022automatic,huang2023contrastive,moummad2022supervised}, and adapt convolutional neural networks (CNNs) on the mel-spectrogram, a synthetic image, for robust respiratory sound classification~\cite{purwins2019deep,kong2020panns,li2022explainable}, \eg, ResNet \cite{nguyen2022lung,ma2019lungbrn,li2021lungattn,gairola2021respirenet,xu2021arsc}, ResNeSt \cite{wang2022domain}, Temporal Convolutional Network \cite{zhao2022automatic} and VGGish-BiGRU \cite{huang2023contrastive}.  

However, existing methods often ignore the differences between spectrograms and natural images, \eg, spectrograms do not contain the visual semantic in natural images~\cite{ren2017regularized,ren2021three}. In particular, the two axes of spectrograms represent time and frequency, which are completely different from the two spatial coordinates in natural images. Consequently, we can treat the two dimensions differently based on the different physical meanings to solve the problem more effectively.
In literature, Vision Transformer (ViT) has been widely used in many applications \cite{dosovitskiy2021an,he2023hierarchical,ding2023vlt, chen2022attention,ji2023dual,gong21b_interspeech,bae2023patch}, which splits an image into a sequence of image patches and extract the attention information from patches using self-attention \cite{dosovitskiy2021an}. 
Inspired by these methods, a Multi-View Spectrogram Transformer (MSVT) is proposed to capture the acoustic characteristics. 


The proposed MSVT is motivated by two factors. 
1) The mel-spectrogram provides fine resolution for lower frequencies but coarser resolution for higher frequencies \cite{ren2016sound}, while the square patches in audio spectrograms \cite{gong21b_interspeech,bae2023patch} often results in limited spectral information of high-frequency patches. 2) The same type of respiratory sounds may contain significant frequency changes~\cite{kong2020panns}, which leads to misalignment of patches in frequency if the patch is too small, or losing the critical frequency information if the patch is too large. 
To address these two challenges, the proposed MVST embeds different views of spectral characteristics into ViTs using a novel patching scheme that treats the time and frequency axes differently. Specifically, the input mel-spectrogram is split into patches of shape varying from $16\times16$ as in the ViT \cite{dosovitskiy2021an} to $256\times1$ by increasing the bandwidth while reducing the time interval.  
This partially solves the issue of frequency misalignment. More importantly, patches of a specific size capture a unique view of the mel-spectrogram, and collectively these multi-view spectrograms capture different spectral views of a given audio signal. 
All the patches are then embedded with positional information and fed into the transformer encoders to perform multi-head self-attention, which deeply exploits the spectral characteristics of the audio signal. A gated fusion scheme \cite{liu2022cross} is then designed to automatically highlight the most suitable view of spectral features for final classification. 

Our main contributions can be summarized as follows. 1)~The proposed MVST better captures the spectral characteristics of audio signals in different views, and automatically highlights the most suitable spectral view through the gated fusion mechanism. 2)~The proposed MVST is evaluated on the ICBHI challenge dataset \cite{rocha2018alpha}. It significantly outperforms state-of-the-art methods for respiratory sound classification. 

\section{Proposed Multi-View Spectrogram Transformer}

\subsection{Overview of Proposed MVST}
The proposed MVST for respiratory sound classification is shown in Fig.~\ref{fig:main-diagram}. The raw audio signal is first converted into a mel-spectrogram using STFT. To capture the multi-view spectral characteristics embedded in different frequency bands at different time intervals, the spectrogram is split into different-sized patches as described in Sec.~\ref{sec:split}. These patches together with the position embeddings are then fed into the transformer encoders, in which the multi-head self-attention (MSA)~\cite{dosovitskiy2021an} is utilized to exploit the audio characteristics residing in patches and the long-range interactions between patches as described in Sec.~\ref{sec:MVTE}. As a result, patches with bare spectral information will be less weighed while patches with rich spectral information will be weighed with a greater attentive importance. The recurrent spectral patterns can be more effectively captured across patches. Lastly, the spectral features of different views are fused through the gated fusion scheme for final classification as described in Sec.~\ref{sec:gfs}. 

Specifically, a raw audio signal $s(t)$ is divided into $M$ overlapping fragments $\{\bm{x}_0, \bm{x}_1, \dots, \bm{x}_{M-1}\}$, where each fragment $\bm{x}_i=\{x_i[n],n=0,1,\dots,N-1\}$. 
The spectrogram $\boldsymbol{S} = \{|f_{i,k}|^2\}$, $i=0,1,\dots,M-1, k=0,1,\dots,N-1$, is formed by applying STFT as,  
\begin{equation}\label{eqn:spectrogram}
	f_{i,k} = \sum_{n=0}^{N-1}x_{i}[n]\exp\left\{-j2\pi \frac{kn}{N}\right\}.
\end{equation}
A linear transformation is then applied on the spectrogram to convert it into mel scale to derive the 
mel-spectrogram. 


\subsection{Multi-View Patch Splitting} \label{sec:split}

 
The spectrogram $\bm{S}$ is first resized to $H\times W$. The multi-view patch splitting then divides it into patches of size $\mu \times \nu$ to obtain the set of patches $\bm{\mathsf{P}} \in \mathbb{R}^{\frac{H}{\mu}\times \frac{W}{\nu}\times \mu\nu}$, which represent $n_T = \frac{H}{\mu}\frac{W}{\nu}$ token vectors of dimension $\mu\nu$. In the standard ViT \cite{dosovitskiy2021an}, a common choice is a square with $\mu=\nu=16$. 
The proposed multi-view patch splitting generates non-overlapped patches of different shapes from the spectrogram by considering the time and frequency differently, \eg, for a patch of size $2^\ell \times \frac{H}{2^\ell}$, $\bm{\mathsf{P}}^{\ell} \in \mathbb{R}^{\frac{H}{2^\ell}\times 2^\ell \times H}$. Each group of patches represent a unique view of audio spectral characteristics, \eg, square patches represent the spectral information in a specific frequency band at a specific time interval while slim and long patches of size $H\times1$ represent the entire frequency response at a specific time instance. In practice, five groups of patches $\{{\bm{\mathsf{P}}}^{\ell},\ell=0\dots 4\}$ are generated from the spectrogram of size $256\times1024$, whose size ranges from $256\times1$, $128\times2$, $64\times4$, $32\times8$ to $16\times16$, as shown in Fig.~\ref{fig:main-diagram}. 
\begin{figure}[ht]
	\centering
	\subfigure[]{
		\includegraphics[height=0.3\linewidth]{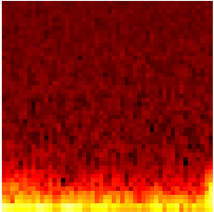}
        \label{fig:seg_a}
	}
 \quad\quad
	\subfigure[]{
		\includegraphics[height=0.3\linewidth]{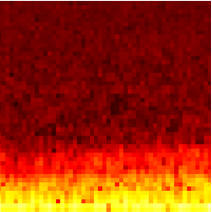}
        \label{fig:seg_b}
	} 
	\caption{Two example spectrograms sliced from the same crackling \& breathing cycle at different time intervals.}
	\label{fig:spectrograms_compare}
\end{figure}

The proposed multi-view patch splitting is motivated by the audio characteristics of respiratory sounds. Fig.~\ref{fig:spectrograms_compare} shows two example spectrograms derived from the segments of the same crackling breathing cycle but at different time intervals. These two spectrograms should look similar but there is a clear frequency shift in critical components, \ie, the frequency of critical components in Fig.~\ref{fig:spectrograms_compare}(b) are higher than that in Fig.~\ref{fig:spectrograms_compare}(a). If the original $16\times16$ patching scheme applies, the audio characteristics in a patch of one spectrogram may not align well to that of another spectrogram, causing misalignment of critical frequency components. In contrast, the newly generated patches have an increasing bandwidth and a reduced time interval, as shown in Fig.~\ref{fig:main-diagram}. These slim and long patches are more tolerant of the frequency shift in respiratory sounds. 


\subsection{Multi-view Transformer Encoder}
\label{sec:MVTE}
As shown in Fig.~\ref{fig:main-diagram}, there are five groups of transformer encoders, one for each view of spectrograms. 
Each transformer encoder takes the $i$-th group of patches ${\bm{\mathsf{P}}}^i$ as the input and embeds them into $d_T$ dimensions using an affine transformation to obtain token vectors $\bm{F}^{i}_0 \in \mathbb{R}^{n_T\times d_T}$ as in \cite{dosovitskiy2021an}. A positional encoding $\bm{P}^i \in \mathbb{R}^{n_T\times d_T}$ is added to $\bm{F}^{i}_0$, \ie, $\tilde{\bm{F}}^{i}_0=\bm{F}^{i}_0+\bm{P}^i$, to supplement the positional information for each token, \ie, the frequency band and the time interval of the audio spectrogram. 
Next, $L$ successive transformer blocks $\mathcal{B}^i_j$ are adopted to extract the attentional features, with each block taking the output of the previous block $\bm{F}^i_{j-1}$ as the input, \ie, $\bm{F}^i_j = \mathcal{B}^i_j(\bm{F}^i_{j-1})$, where $\bm{F}^i_j \in \mathbb{R}^{n_T\times d_T}$. The input to the first block is $\tilde{\bm{F}}^{i}_0$ and the output of the last block $\bm{F}^i_L$ is taken as the final feature map. 

For each transformer block $\mathcal{B}^i_j$, as shown in Fig.~\ref{fig:main-diagram}, a multi-head self attention (MSA) \cite{dosovitskiy2021an} is utilized, followed by a multi-layer perceptron (MLP), where the MSA computes the global relations between patch tokens 
as in \cite{dosovitskiy2021an}, 
\begin{equation}
    {{\bm{Z}}^i_j} = \mathcal{F}_{MSA}(\mathcal{F}_{LN}(\bm{F}^i_{j-1})) + \bm{F}^i_{j-1}, \quad j=1\dots L,
\end{equation}
where $\mathcal{F}_{LN}$ represents the layer normalization \cite{ba2016layer}. Then, the MLP with a GELU (Gaussian Error Linear Unit) activiation function $\epsilon$ derives the output of the $j$-th block as in \cite{dosovitskiy2021an},  
\begin{equation}
{\bm{F}}^i_j = \epsilon(\mathcal{F}_{MLP}(\mathcal{F}_{LN}({\bm{Z}}^i_j))) + {\bm{Z}}^i_j, \quad j=1\dots L.
\end{equation}

\subsection{Gated Fusion Scheme}
\label{sec:gfs}
In our specific task, five groups of different-sized patches generate five groups of feature maps $\{{\bm{F}}^i_L\}_{i=0}^4$ through the transformer encoders, each representing a unique view of acoustic characteristics. To constitute a complete view of audio spectrograms from various aspects, the multi-view features are fused using the gated fusion scheme \cite{liu2022cross}, which weighs the feature map $\bm{F}^i_L$ using the gated coefficients $\bm{G}^i \in \mathbb{R}^{n_T\times d_T}$,  
\begin{equation}
    \tilde{\bm{F}} = \sum_{i=0}^4 \bm{G}^i \odot \bm{F}^i_L, 
\end{equation}
where $\odot$ indicates the Hadamard (element-wise) product. The gated coefficients are derived as, 
\begin{equation}
    \bm{G}^{i}=\sigma(\sum_{i=0}^4 \bm{W}^i \bm{F}^{i}_L),
\end{equation}
where $\bm{W}^i \in \mathbb{R}^{d_T\times d_T}$ is the learnable weight matrices and $\sigma$ is the sigmoid function. The final prediction for a spectrogram is derived using a MLP as $\hat{y} = \mathcal{F}_{MLP}(\tilde{\bm{F}})$. This gated fusion scheme could automatically highlight the most suitable view of spectral characteristics for respiratory sound classification.   

\section{Experimental Results}
\subsection{Experimental Settings}
The proposed MVST is evaluated on the largest publicly available respiratory sound dataset, the ICBHI dataset \cite{rocha2018alpha}, which contains 6,898 respiratory cycles, including 3,642 normal breathing, 1,864 crackling breathing, 886 wheezing breathing and 506 breathing with both crackling \& wheezing. The sampling rate is 4kHz, 10kHz or 44.1kHz, with a recording duration between 10 and 90 seconds. Each recording includes an average of seven breathing cycles. 
Following the official experimental settings in~\cite{rocha2018alpha,nguyen2022lung}, 60\% of the patient recordings are randomly selected for training and the rest 40\% for testing. 
Standard evaluation metrics \cite{rocha2018alpha} such as specificity (\textit{SP}), sensitivity (\textit{SE}) and average score (\textit{AS}) are used. 
For the proposed MVST, the cross-entropy loss is used with a learning rate of $5 \times 10^{-5}$. The AdamW optimizer is used with a fixed weight decay of $1 \times 10^{-6}$. To ensure the convergence and the efficiency of the model, the maximum number of epochs is set to $50$ and the batch size is $8$. 





\subsection{Ablation Study}
An ablation study is carried out to evaluate the effectiveness of the proposed MVST. For the five compared single-view methods, the spectrograms are divided into patches of the specific size and fed into the ViT for classification. 
The results are summarized in Table~\ref{tab:abla_results}. 
It can be observed that: 
1) Many other views perform competitively compared with $16\times16$ baseline method, which motivates us to jointly learn from different views of different discrimative features. 
2) The $128\times2$ patching scheme produces the highest AS of 62.94\% among single-view methods, which demonstrates the effectiveness of splitting into slim and long  patches by reducing the time interval while increasing the bandwidth, as this design partially alleviates the problem of frequency misalignment in respiratory sounds. 
3) The proposed MVST produces significantly better \textit{SP}, \textit{SE} and \textit{AS} than all single-view methods, which is attributed to the multi-view spectrograms in effectively capturing the audio characteristics, and the gated fusion scheme in combining the multi-view spectrogram features for robust respiratory classification. 

\begin{table}[htpb]
	\centering
	\caption{Ablation study of the proposed MVST. 
 }
		\begin{tabular}{lccc}
			\toprule
		  Patch size & \textit{SP}(\%) & \textit{SE}(\%) & \textit{AS}(\%)\\
            \midrule
            $16\times16$ (Baseline) & 78.48 & 46.62 & 62.55 \\
            $32\times8$ & 
            79.31 & 42.61 & 60.96 \\
            $64\times4$ & 
            84.55 & 39.68 & 62.11 \\
            $128\times2$ & 
            80.17 & 45.71 & 62.94 \\
            $256\times1$ & 
            84.63 & 40.14 & 62.38 \\ \midrule
            \textbf{Proposed MVST} & \textbf{81.99} & \textbf{51.10} & \textbf{66.55} \\
            \bottomrule
		\end{tabular}
	\label{tab:abla_results}
\end{table}

\subsection{Comparison to State-of-the-art Methods}
Many deep neural networks have been utilized as the backbone for respiratory sound classification, \eg, 
bi-ResNet in \textbf{LungBRN}~\cite{ma2019lungbrn}, ResNet-Att in \textbf{LungAttn}~\cite{li2021lungattn}, \textbf{ResNeSt}~\cite{wang2022domain}, 
ResNet-34 in \textbf{RespireNet}~\cite{gairola2021respirenet}, 
bi-ResNet-Att in \textbf{ARSC-Net}~\cite{xu2021arsc}, \textbf{CNN6}~\cite{moummad2022supervised}, 
\textbf{Co-tuned ResNet-50}~\cite{nguyen2022lung}
and audio spectrogram transformer with patch-mix and contrastive learning (\textbf{AST + patch-mix CL})~\cite{bae2023patch}. 
The comparison with these state-of-the-art methods are summarized in Table~\ref{tab:sota_results}.\footnote{The results of all the compared methods are obtained from \href{https://paperswithcode.com/sota/audio-classification-on-icbhi-respiratory}{\underline{here}}.}
It can be observed that: 1) Many models utilize  traditional CNN models for respiratory classification~\cite{nguyen2022lung,ma2019lungbrn,li2021lungattn,gairola2021respirenet,xu2021arsc,wang2022domain,moummad2022supervised}, while the most recent work \cite{bae2023patch} utilizes audio spectrogram transformer \cite{gong21b_interspeech} and achieves the previous best results on the ICBHI dataset. This partially motivates us to utilize the ViT as the backbone network. 
2)~The proposed MVST significantly outperforms all the compared methods in terms of all three criteria \textit{SP}, \textit{SE} and \textit{AS}. Specifically, it outperforms the second-best approach AST + patch-mix CL \cite{bae2023patch} by 0.33\%, 8.09\% and 4.18\% in terms of \textit{SP}, \textit{SE} and \textit{AS} respectively. 
Compared to Co-tuned ResNet-50~\cite{nguyen2022lung}, the proposed MVST achieves a performance gain as high as 8.26\% in terms of the key metric \textit{AS}. 

\begin{table}[htpb]
	\centering
	\caption{Comparisons with state-of-the-art methods.} 
		\begin{tabular}{lccc}
			\toprule
			Methods  & \textit{SP}(\%) & \textit{SE}(\%) & \textit{AS}(\%)\\
            \midrule
            LungBRN \cite{ma2019lungbrn} & 69.20 & 31.10 & 50.16 \\
            LungAttn \cite{li2021lungattn} & 71.44 & 36.36 & 53.90 \\
            ResNeSt \cite{wang2022domain} & 70.40 & 40.20 & 55.30 \\
            RespireNet \cite{gairola2021respirenet} & 72.30 & 40.10 & 56.20 \\
            ARSC-Net \cite{xu2021arsc} & 67.13 & {46.38} & 56.76 \\
            CNN6 \cite{moummad2022supervised} & 75.95 & 39.15 & 57.55\\
            Co-tuned ResNet-50 \cite{nguyen2022lung} & 79.34 & 37.24 & 58.29 \\
            AST + patch-mix CL \cite{bae2023patch} & {81.66} & 43.01 & {62.37} \\ \midrule
            \textbf{Proposed MVST} & \textbf{81.99} & \textbf{51.10} & \textbf{66.55} \\
            \bottomrule
		\end{tabular}
	\label{tab:sota_results}
\end{table}

The performance gain of the proposed MVST is attributed to two major factors. 1) The ViT backbone could better extract spectrogram features. Specifically, the ViT partitions the spectrograms into patches with explainable physical meanings, which correspond to the frequency response in a specific frequency band at a particular time interval. The subsequent attention blocks then could utilize the MSA mechanism to better extract the global interactions among patches. 2)~The proposed MVST addresses the challenges of frequency misalignment in spectrograms by splitting an audio spectrogram into different-sized patches, capturing the spectral characteristics from different views, and alleviating the problem of frequency shift in respiratory sounds. The ablation study shows that using different patch sizes is beneficial for classifying respiratory sounds, and the gated fusion integrates the discriminative information extracted from different views, leading to the superior performance of the proposed MVST.

\section{Conclusion}
In this paper, a Multi-View Spectrogram Transformer is proposed to deal with the challenges of classifying respiratory sounds. Different from natural image, we treat the two axes of the synthetic image, mel-spectrogram, differently based on their different physical meanings. The proposed multi-view patch splitting scheme captures different views of the frequency response of audio signals, better comprehends the acoustic elements of respiratory sounds and overcomes the problem of potential frequency shifts. The transformer encoders better extract the spectral characteristics within each patches and the global interaction among patches. 
These multi-view features are then effectively aggregated by the gated fusion scheme to automatically highlight the most discriminant features in a specific scenario. Experimental results show that the proposed MVST significantly outperforms all the state-of-the-art methods on the ICBHI dataset. 

\clearpage
\balance
\small{
\bibliographystyle{IEEEbib}
\bibliography{refs}
}

\end{document}